\documentclass[pre,aps,twocolumn]{revtex4}
\usepackage[pdftex]{graphicx}
\usepackage{subfigure,amsbsy,amsmath}

\usepackage[latin1]{inputenc}

\vfil

\usepackage{amssymb}
\usepackage{amsfonts}
\usepackage{mathrsfs}

\begin{document}

\title{Vortex Dynamics in Cerebral Aneurysms}

\author{Greg Byrne$^1$ and Juan Cebral$^2$}

\affiliation{$1$Center for Nonlinear Science, Georgia Institute of Technology, Atlanta, GA 30332, USA}
\affiliation{$2$Center for Computational Fluid Dynamics, George Mason University, Fairfax, VA 22030, USA}

\begin{abstract}
We use an autonomous three-dimensional dynamical system to study embedded vortex structures that are observed to form in computational fluid dynamic simulations of patient-specific cerebral aneurysm geometries.  These structures, described by a vortex which is enclosed within a larger vortex flowing in the opposite direction, are created and destroyed in phase space as fixed points undergo saddle-node bifurcations along vortex core lines.  We illustrate how saddle-node bifurcations along vortex core lines also govern the formation and evolution of embedded vortices in cerebral aneurysms under variable inflow rates during the cardiac cycle. 

 
\end{abstract}

\pacs{PACS numbers:}
\maketitle

\section{Introduction}
\label{sec:introduction}

Computational fluid dynamic simulations of patient-specific vascular geometries are providing new insights into the connections between hemodynamics and the initiation, progression and treatment of cerebral aneurysms--balloon like dilations that occur along arterial walls as a result of abnormal wall shear stresses \cite{CebralCAPMF05,cebral_characterization_2005,kadirvel_influence_2007,penn_hemodynamic_2011,chalouhi_biology_2012,sforza_hemodynamics_2009}.  Recent studies linking aneurysm rupture to the formation of vortices motivate the need for a more fundamental understanding of swirling blood flow patterns and their evolution during the cardiac cycle \cite{cebral_association_2011}.  One-dimensional vortex skeletons called vortex core lines are being used to facilitate this process and advance our knowledge of vortices within cerebral aneurysms \cite{perry_aspects_1984,byrne_quantifying_2013}.

When a vortex in the free-slip region of a cerebral aneurysm makes contact with the aneurysm wall, the end of its vortex core forms a stagnation point that, much like the eye of a hurricane, organizes spiral shear line patterns on the no-slip domain \cite{goubergrits_statistical_2011}.  These types of stagnation points have been studied mathematically using dynamical systems derived from truncated Taylor series expansions of the wall shear stress \cite{gambaruto_flow_2012}.

In this work, we use a three-dimensional autonomous dynamical system to investigate the formation of intra-aneursym blood flow structures in which a vortex is enclosed within a larger vortex flowing in the opposite direction.  These embedded vortices are often observed to form in computational fluid dynamic simulations run on an extensive database of patient-specific aneurysm geometries.  We focus specifically on the formation of stagnation points and seek to describe their role in controlling the large-scale structure of swirling flow along vortex core lines.

In Sec. \ref{sec:Background} we provide some background by describing the computational fluid dynamic simulations that are used to study the hemodynamics of patient-specific cerebral aneurysm geometries.  We describe the construction of vortex core lines and the use streamlines to visualize the vortices that surround them.  We also provide an example of an embedded vortex that forms in a cerebral aneurysm from our database.  In Sec. \ref{sec:PhaseSpace} we construct a dynamical system to recreate this embedded vortex structure in phase space.  A single control parameter is varied to investigate how the flow dynamics change as saddle-node bifurcations are induced along a vortex core line.  In Sect. \ref{sec:CFDSpace} we identify stagnation points along the vortex core lines of our hemodynamic simulations and show how saddle-node bifurcations have the same effect on the blood flow dynamics as inflow rates are changed during the cardiac cycle.  Section \ref{sec:Conclusions} summarizes the work and provides some conclusions.  

\section{Background}
\label{sec:Background}

\subsection{Patient-Specific Hemodynamic Simulations of Cerebral Aneurysms}

A total of 210 patient-specific vascular models containing cerebral aneurysms were constructed from medical images using three-dimensional rotational angiography (3DRA).  Unstructured meshes consisting of between 2-5 million tetrahedral elements were generated in these geometries to run computational fluid dynamic simulations.  Pulsatile inflow conditions with a Womersley velocity profile were imposed at the inlets and a no-slip condition was enforced along vessel walls which were assumed to be rigid.  The blood was considered to be incompressible and Newtonian.  Because of these assumptions, the unsteady three-dimensional Navier-Stokes and the continuity equation were solved using a finite element solver.  

Two cardiac cycles were run with a time step of 0.01 seconds using a heart rate of 60 beats per minute.  Each cardiac cycle was discretized into $N=100$ samples.  Measurements of the velocity vector field were recorded at the element nodes during the second cycle, resulting in a data ensemble of 100 snapshots: $u_n({\bf x},t_n)$, $n=1\dots100$.  Additional details about the computational pipeline can be found in \cite{CebralCAPMF05}.

\subsection{Construction of Vortex Core Lines}

Vortex core lines are constructed in the aneurysm region by identifying the locus of points that satisfy the following two conditions \cite{peikert_parallel_1999}: (1) the acceleration at a point is proportional to the velocity at that point, and (2) the velocity gradient tensor produces a pair of complex conjugate eigenvalues $\lambda_{C}$ and one real eigenvalue $\lambda_{R}$.  The first condition can be expressed mathematically by an eigenvalue equation 

\begin{equation}
J {\bf u}=\lambda {\bf u} 
\label{eq:eigen}
\end{equation}

where the Jacobian $J=\nabla {\bf u}$.  Equation (\ref{eq:eigen}) is satisfied when the velocity vector is an eigenvector of the Jacobian.  

The second condition ensures that flow undergoes swirling in the neighborhood of the vortex core line.  Swirling occurs in a plane spanned by the eigenvectors ${\bf \xi}_{C}$ corresponding to the complex conjugate eigenvalues $\lambda_{C}$.  When transported along the eigendirection ${\bf \xi}_{R}$ associated with the real eigenvalue $\lambda_{R}$, this swirling flow creates the tornado-like motion of a vortex.  
 
Each tetrahedral element in the unstructured mesh is checked to see if it satisfies equation (\ref{eq:eigen}).  This is done using the reduced velocity method of Sujudi and Haimes \cite{sujudi_identification_1995}.  The first step is to use the element's linear interpolation coefficients (known as shape functions) to compute the Jacobian.  The Jacobian is then diagonalized, and processing continues only if the eigenvalues contain a complex conjugate eigenvalue pair.  Reduced velocities ${\bf w}={\bf u}-\left( {\bf u} \cdot {\bf \xi}_R \right) {\bf \xi}_R$ are formed in each of the nodes by subtracting the velocity component in the direction of the real eigenvector corresponding to $\lambda_{R}$.  

Linear interpolation is used on each of the element faces to locate points where the reduced velocity field is equal to zero.  Zeros of the reduced velocity field corresponds to centers of swirling flow on the element faces.  If two faces are found to contain a zero, a line segment approximating the location of the vortex core through the element is constructed between them.  

Small gaps are known to form between core line segments in neighboring elements because the Jacobian is constant in a tetrahedral element and piece-wise linear across the computational domain.  When needed, mesh refinement can be used to close these gaps and resolve vortices on a smaller scale.  

\subsection{Embedded Vortex Structures}
\label{sec:EmbeddedAneurysm}

\begin{figure}[htbp]   
  \center
  \includegraphics[height=17cm,width=\columnwidth]{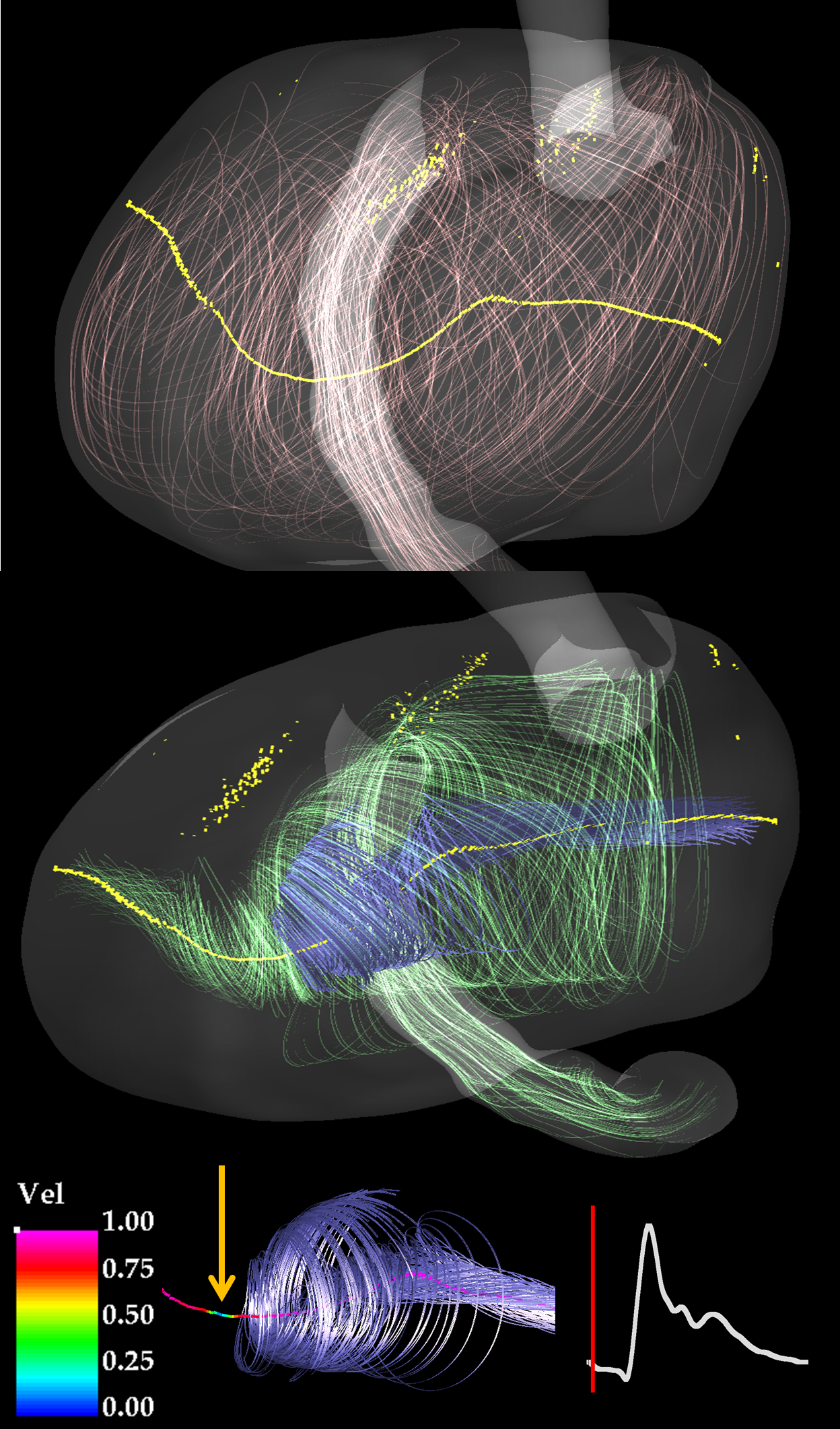}
   \caption{(top) Streamlines generated by computational fluid dynamic simulations are used to visualize the global flow patterns inside a patient-specific aneurysm geometry.  The vortex core lines shown in yellow provide important information about the organization of swirling blood flow within an aneurysm.  (middle) Streamlines initialized at either end of the vortex core line reveal the formation of an embedded vortex structure.  (bottom) Stagnation points are identified by looking for zeros of the rescaled velocity magnitude along the vortex core line.  The yellow arrow indicates the location of a stagnation point that forms as the inner vortex meets the outer vortex.  The cardiac wave form on the right shows the location in the cardiac cycle where the embedded vortex was formed.}
\label{fig:figure1}   
\end{figure}

Streamlines are used to visualize the flow patterns within the aneurysms.  When initialized in the parent vessel, the streamlines fill the aneurysm volume and illuminate the global flow pattern within it.  If the flow pattern is complex, the streamlines do not always fully resolve its spatial structure.  An example in which this occurs is shown in Fig. \ref{fig:figure1} (top).  Adding more streamlines can result in dense ``spaghetti plots'' that make the task even more difficult.

The vortex core lines in the plot, shown in yellow, help identify the underlying spatial structure by providing an axis around which the flow swirls.  The long horizontal core line shown in the figure captures the large recirculation zone that forms within the entire aneurysm volume.  The fuzzier, two-dimensional structures observed above the main core line are vortex sheets that form in a shear layer.  

Two sets of streamlines are initialized near the core line to better resolve the swirling flow.  They are shown in Fig. \ref{fig:figure1} (middle).  The first set, colored in blue, start on the right side of the core line and travel towards the left until they suddently begin to decelerate and spiral outward.  The second set, colored in green, start on the left side of the core line and travel towards the right until they meet and envelop the first set of streamlines.  This embedded flow structure occurs in many of the cases from our aneurysm database and is always associated with the presence of stagnation points along the vortex core line.  

An example of a stagnation point is shown in Fig. \ref{fig:figure1} (bottom).  Its location, marked by the yellow arrow, is where the interior vortex meets the exterior vortex.  The velocity magnitude along the core line has been rescaled between zero and one to simplify the comparison between different aneurysm cases.  

The cardiac wave form on the right shows where in the cardiac cycle the embedded vortex is observed to form.  Embedded vortices in other aneurysm geometries from our database showed variability in their location on the inflow curve, suggesting that there is no preferred location in the cardiac cycle where they form.

\section{The Embedded Vortex Model}
\label{sec:PhaseSpace}

A jerk system \cite{sprott_simple_1997,sprott_chaos_2003,sprott_elegant_2010} of the form $\frac{d^3x}{dt^3}=f\left(\frac{d^2x}{dt^2},\frac{dx}{dt},x \right)$ is used to recreate the embedded vortex structure described in Sec. \ref{sec:EmbeddedAneurysm}.  We investigate embedded vortices in the phase space of dynamical systems because fixed points \cite{henri_poincare_sur_1886,henri_poincare_sur_1885,henri_poincare_sur_1882,henri_poincare_sur_1881,perko_differential_2001,ott_chaos_2002}, bifurcations \cite{arnold_singularities_1985,arnold_catastrophe_1992,gilmore_catastrophe_1993}, the relationship between fixed points and vortex core lines \cite{byrne_connecting_2013}, and the large-scale structural reorganization of flows under control parameter variation \cite{letellier_large-scale_2005} have all been extensively studied and are all well understood.  


The third-order equation is expanded into a set of three coupled first-order ordinary differential equations of the form 

\begin{equation}
  \begin{array}{l}
    \dot{x} = y \\
    \dot{y} = z \\
    \dot{z} = F(x,y,z;c);
  \end{array}
\label{eq:3dode}
\end{equation}

where each phase space coordinate except the first is the time derivative of the previous.  It is convenient to assume that the source function $F(x,y,z;c)$ can be separated into a sum of two functions $F(x,y,z;c)=G_1(x;c_1)+G_2(y,z;c_2)$, since the fixed points are then determined by the condition $G_1(x;c_1)=0$.  Setting $G_1(x;R)=x^2-R$ allows us to control the number, location and stability of the fixed points by varying the parameter $R$.  A maximum of two fixed points, located at ${\bf x}_{f_1}=(-\sqrt{R},0,0)$ and ${\bf x}_{f_2}=(\sqrt{R},0,0)$, can exist along the $x$ axis.  These fixed points are created and destroyed in a saddle-node bifurcation as the control parameter $R$ is varied.    

We set $G_2(y,z;A,B)=Ay+Bz^2$ and leverage catastrophe theory \cite{gilmore_catastrophe_1993,poston_catastrophe_1996} to determine suitable values for the control parameters $A$ and $R$.  This is done by noticing that when evaluated at any fixed point ${\bf x}_f$, the Jacobian produces a characteristic polynomial

\begin{equation}
J = \left[  \begin{array}{ccc}
0 & 1 & 0 \\
0 & 0 & 1 \\
2x_{f} & A & 0 \end{array} \right] 
\stackrel{\rm char.~poly}{\longrightarrow}
\lambda^3 - A \lambda - 2x_f 
\label{eq:cp}
\end{equation}

whose roots determine the stability properties (focus or saddle) 
of the fixed point.  The substitution $(a,b)=(-A,-2x_{f})$ transforms 
(\ref{eq:cp}) into the canonical form of the cusp catastrophe 
$A_3'=\lambda^3 +a \lambda + b$ \cite{gilmore_catastrophe_1993}.  The 
resulting catastrophe set partitions the control parameter plane into two 
subsets, the first of which guarantees that the fixed point is a saddle 
and a second which guarantees that the fixed point is a focus.  The catastrophe 
set is used to select a value of $A=-5$ such that the fixed points retain focal 
stability as they undergo a saddle-node bifurcation for values of $R=[-1,1]$.

\begin{figure*} 
\begin{minipage}{2.0\columnwidth}
\centering
\setlength\fboxsep{0pt}
\setlength\fboxrule{0.25pt}
\includegraphics[height=15cm,width=0.95\columnwidth]{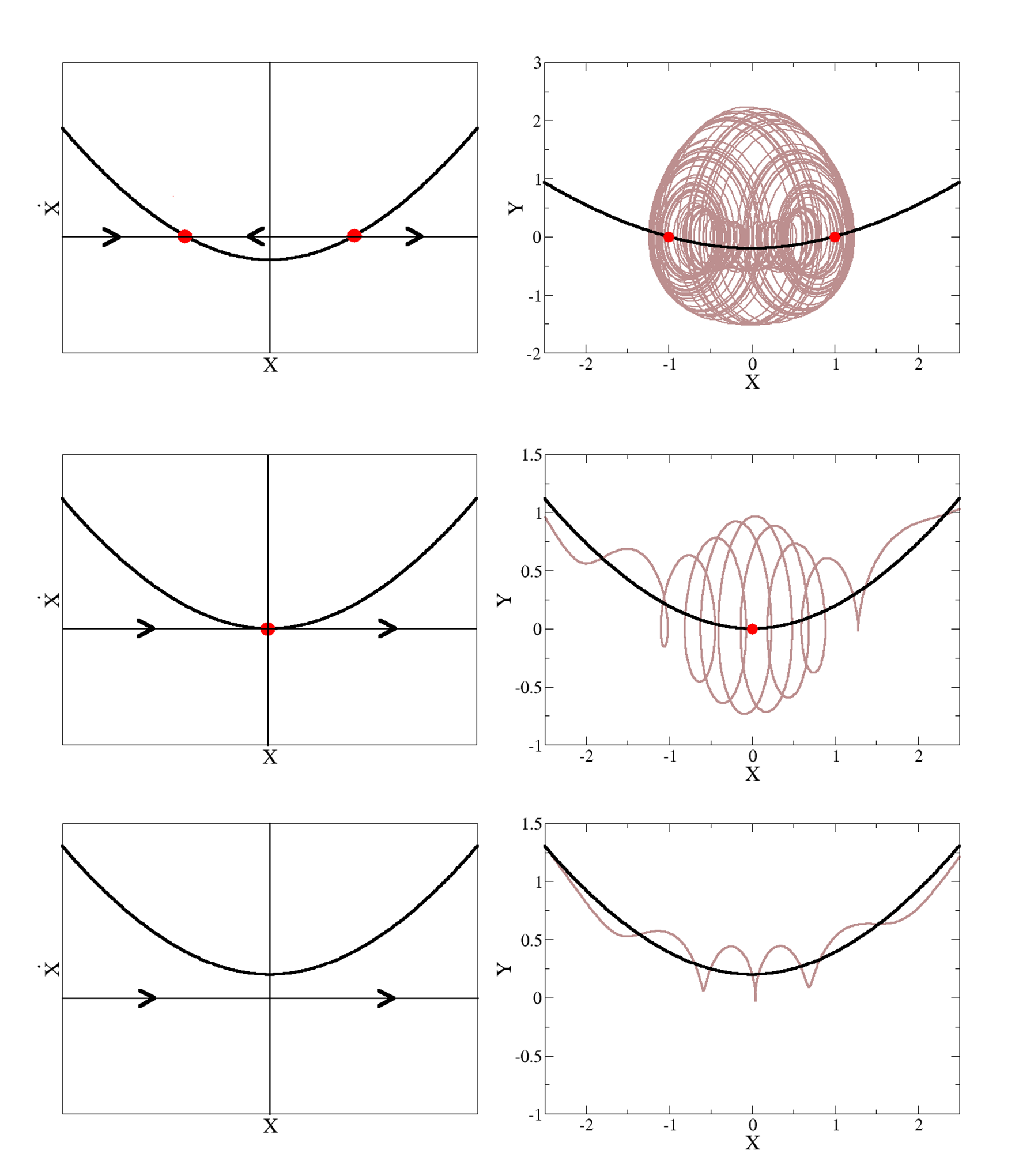}
\caption{The phase space dynamics are described for the embedded vortex model in (\ref{eq:3dode}) using control parameter values $(A,B)=(-5,0.2)$ and varying $R$.  Fixed points are shown in red, vortex core lines in black and phase space trajectories in brown.  The arrows along the x-axis in the left column indicate the flow direction along the vortex core line.  The right column shows the evolution of phase space trajectories along the core line.  (top row) For $R=1$, the alternating stability of the two fixed points allows the swirling flow to be recirculated in a way that creates an embedded vortex structure.  (middle row) For $R=0$, the fixed points collide and the flow becomes uni-directional along the core line.  (bottom row) For $R=-1$, the fixed point disappears and the flow along the core line remains uni-directional.}
\label{fig:figure2}  
\end{minipage}
\end{figure*}

Swirling phase space trajectories are identified using connecting 
curves--one-dimensional sets whose properties are analogous to the vortex core 
lines defined for hydrodynamic flows \cite{gilmore_connecting_2010}.  
Since the dynamical system is three-dimensional, subsets of the 
connecting curve around which the flow undergoes a swirling motion 
(referred to as vortex core lines for the remainder of this work) are 
known to pass through fixed points with focal 
stability \cite{byrne_connecting_2013}.

When $R>0$, fixed points along the vortex core line experience an alternation of 
stability (stable focus - unstable outset) with (unstable focus - stable inset) 
along the $x$ axis.  This is determined by examining the properties 
of $G(x,R)=x^2-R$.  If $G' > 0$ at a fixed point then one eigenvalue 
is positive and the other two are negative, or have
negative real part.  If $G' < 0$ the reverse is true:
one eigenvalue is negative and the other two are either
positive or have positive real part.  

In cases where there are more 
than two fixed points, interior fixed points that are stable foci with 
unstable outsets send the flow to neighboring
fixed points on its left and right, which are unstable
foci with stable insets.  This situation allows for multiple inner vortices 
to be embedded within the same outer vortex.  

Figure \ref{fig:figure2} shows the phase portraits for (\ref{eq:3dode}) using $F(x,y,z;c)=x^2-R+Ay+Bz^2$ and $(A,B)=(-5,0.2)$.  Fixed points are shown in red, vortex core lines in black and phase space trajectories in brown.  The flow direction along the vortex core line, indicated by the arrows on the x-axis, are shown in the column on the left.  The behavior of the phase space trajectories, initialized near the vortex core lines, is shown in the column on the right.  Each row represents a different value of $R$ to illustrate how the saddle-node bifurcation controls the large-scale structure of the flow along the vortex core line.  From top to bottom, these value are $R=[1,0,-1]$.

In Fig. \ref{fig:figure2}(top), $R=1$ and two fixed points are located along the vortex core line at ${\bf x}_{f_1}=(-1,0,0)$ and ${\bf x}_{f_2}=(1,0,0)$.  A bounded attractor forms between the fixed points whose alternating stability is responsible for the formation of the embedded vortex.  In Fig. \ref{fig:figure2}(middle), $R=0$ and the fixed point pair collide to form a single degenerate saddle-node type fixed point at the origin.  Trajectories flow along the vortex core line towards the right, slowing down and spiraling away from the fixed point before continuing towards the attractor at infinity.  In Fig. \ref{fig:figure2}(bottom), $R=-1$ and no fixed points exist along the vortex core line.  Unbounded trajectories continue to flow along the core line towards the attractor at infinity.      

\begin{figure*} 
\begin{minipage}{2.0\columnwidth}
\centering
\setlength\fboxsep{0pt}
\setlength\fboxrule{0.25pt}
\includegraphics[height=13cm,width=0.95\columnwidth]{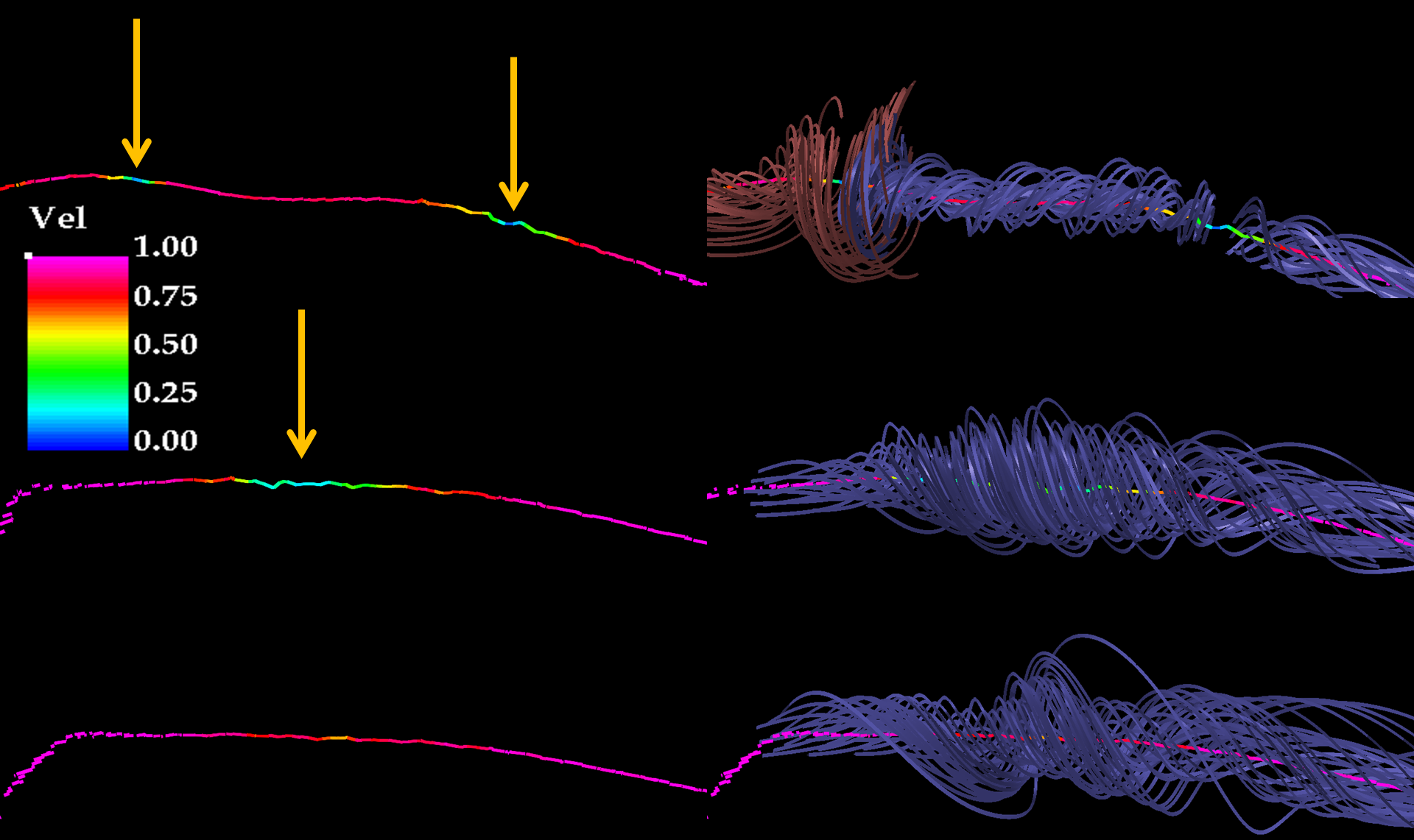}
\caption{The sequence of three successive snapshots(top to bottom row) show how saddle-node bifurcations take place along vortex core lines in hemodynamic simulations of cerebral aneruysms during the cardiac cycle.  The column on the left uses yellow arrows to highlight the locations of stagnation points along the vortex core line.  The column on the right uses streamlines to determine the stability of the stagantion points.  The bifurcation sequence is qualitatively the same as the one shown in Fig. \ref{fig:figure2}.}
\label{fig:figure3}  
\end{minipage}
\end{figure*}

\section{Saddle-node Bifurcations in Cerebral Aneurysms}
\label{sec:CFDSpace}

To apply the results from Sec. \ref{sec:PhaseSpace} to our hemodynamic simulations, we performed a search for saddle-node bifurcations along the vortex core lines.  This was done by animating the core lines and their stagnation points over the cardiac cycle.  An example of a saddle-node bifurcation is shown in Fig. \ref{fig:figure3}.  

The figure shows a sequence of three successive snapshots (the rows from top to bottom) which contain the same segment of vortex core line.  The left column shows how the stagnation points, indicated by the yellow arrows, slide along the length of the vortex core line, collide, and disappear in a saddle-node bifurcation.  The right column uses streamlines to determine the stability of the stagnation points.  The bifurcation sequence shown here is qualitatively the same as that shown in Fig. \ref{fig:figure2}.

The streamlines in the top row indicate that the stagnation points have alternating stability; the one on the left is an unstable focus with stable insets, and the one on the right is a stable focus with unstable insets.  Blue streamlines spiral away from the unstable stagnation point on the right and collide with the red streamlines flowing towards the stable stagnation point on the left.  

The single set of streamlines in the middle row indicate that the flow is uni-directional along the core line (towards the right) after the stagnation points have merged.  This destroys the embedded vortex structure along the core line.  The streamlines shown in the bottom row continue to flow uni-directionally along the vortex core line as the single stagnation point disappears.     

\section{Conclusions}
\label{sec:Conclusions}

This work was motivated by an attempt to understand the large-scale structure of swirling flows organized by vortex core lines in a database of 210 patient-specific aneurysm geometries.  Two types of swirling flow structures were observed to form during the cardiac cycle.  The first involved vortices with a uni-directional flow and the second involved a pair of vortices, the first of which was embedded within the second.  To better understand the mechanisms generating these vortices, a three-dimensional dynamical system was used to study swirling flow structures in phase space.  Catastrophe theory was used to select control parameter values that would produce fixed points along the vortex core line with focal stability.  A single bifurcation parameter $R$ was then varied to force the creation of fixed point pairs along the core line in a saddle-node bifurcation.  A bounded attractor whose flow formed an embedded vortex structure was observed to form in the presence of fixed point pairs with alternating stability.  Uni-directional flow was observed to take place along the core line when the fixed point pair collapsed into a single degenerate fixed point and disappeared.  

To see if the concepts from phase space could be applied to our hemodynamic simulations, a search was performed to identify stagnation points along vortex core lines constructed in the aneurysm database.  Stagnation points were observed to form at the ends of vortex core lines when they came into contact into contact with aneurysm walls.  They were also observed to form along the vortex core lines in the free-slip region.  

Animations over the cardiac cycle revealed that stagnation points in cerebral aneurysms undergo saddle node bifurcations along vortex core lines, and that these bifurcations are responsible for the creation and destruction of embedded vortices in the blood flow.  A more fundamental understanding of vortex dynamics in cerebral aneurysms could lead to improved clinical risk assessment, accelerated thrombosis to promote healing and control algorithms designed to suppress the creation of high risk blood flow patterns.    

\bibliographystyle{unsrt}
\bibliography{references}{}

\begin{thebibliography}{10}

\bibitem{CebralCAPMF05}
Juan~R. Cebral, Marcelo~Adrian Castro, Sunil Appanaboyina, Christopher~M.
  Putman, Daniel Millan, and Alejandro~F. Frangi.
\newblock Efficient pipeline for image-based patient-specific analysis of
  cerebral aneurysm hemodynamics: technique and sensitivity.
\newblock {\em IEEE Trans. Med. Imaging}, 24(4):457--467, 2005.

\bibitem{cebral_characterization_2005}
Juan~R Cebral, Marcelo~A Castro, James~E Burgess, Richard~S Pergolizzi,
  Michael~J Sheridan, and Christopher~M Putman.
\newblock Characterization of cerebral aneurysms for assessing risk of rupture
  by using patient-specific computational hemodynamics models.
\newblock {\em {AJNR} Am J Neuroradiol}, 26(10):2550--2559, December 2005.
\newblock {PMID:} 16286400.

\bibitem{kadirvel_influence_2007}
Ramanathan Kadirvel, Yong-Hong Ding, Daying Dai, Hasballah Zakaria, Anne~M
  Robertson, Mark~A Danielson, Debra~A Lewis, Harry~J Cloft, and David~F
  Kallmes.
\newblock The influence of hemodynamic forces on biomarkers in the walls of
  elastase-induced aneurysms in rabbits.
\newblock {\em Neuroradiology}, 49(12):1041--1053, December 2007.
\newblock {PMID:} 17882410.

\bibitem{penn_hemodynamic_2011}
David~L Penn, Ricardo~J Komotar, and E~Sander~Connolly.
\newblock Hemodynamic mechanisms underlying cerebral aneurysm pathogenesis.
\newblock {\em J Clin Neurosci}, 18(11):1435--1438, November 2011.
\newblock {PMID:} 21917457.

\bibitem{chalouhi_biology_2012}
Nohra Chalouhi, Muhammad~S. Ali, Pascal~M. Jabbour, Stavropoula~I. Tjoumakaris,
  L.~Fernando Gonzalez, Robert~H. Rosenwasser, Walter~J. Koch, and Aaron~S.
  Dumont.
\newblock Biology of intracranial aneurysms: role of inflammation.
\newblock {\em J Cereb Blood Flow Metab}, 32(9):1659--1676, September 2012.

\bibitem{sforza_hemodynamics_2009}
Daniel~M. Sforza, Christopher~M. Putman, and Juan~Raul Cebral.
\newblock Hemodynamics of cerebral aneurysms.
\newblock {\em Ann. Rev. Fluid Mechanics}, 41(1):91--107, 2009.

\bibitem{cebral_association_2011}
J~R Cebral, F~Mut, J~Weir, and C~M Putman.
\newblock Association of hemodynamic characteristics and cerebral aneurysm
  rupture.
\newblock {\em {AJNR.} American journal of neuroradiology}, 32(2):264--270,
  February 2011.
\newblock {PMID:} 21051508.

\bibitem{perry_aspects_1984}
A.~E. Perry and H.~Hornung.
\newblock Some aspects of three-dimensional separation. {II} - vortex
  skeletons.
\newblock {\em Zeitschrift fur Flugwissenschaften und Weltraumforschung},
  8:155--160, June 1984.

\bibitem{byrne_quantifying_2013}
G~Byrne, F~Mut, and J~Cebral.
\newblock Quantifying the large-scale hemodynamics of intracranial aneurysms.
\newblock {\em {AJNR.} American journal of neuroradiology}, August 2013.
\newblock {PMID:} 23928142.

\bibitem{goubergrits_statistical_2011}
L.~Goubergrits, J.~Schaller, U.~Kertzscher, N.~van~den Bruck, K.~Poethkow,
  Ch~Petz, H.-Ch Hege, and A.~Spuler.
\newblock Statistical wall shear stress maps of ruptured and unruptured middle
  cerebral artery aneurysms.
\newblock {\em Journal of The Royal Society Interface}, September 2011.
\newblock {PMID:} 21957117.

\bibitem{gambaruto_flow_2012}
{A.M.} Gambaruto and {A.J.} João.
\newblock Flow structures in cerebral aneurysms.
\newblock {\em Computers \& Fluids}, 65:56--65, July 2012.

\bibitem{peikert_parallel_1999}
R.~Peikert and M.~Roth.
\newblock The {"Parallel} vectors" operator-a vector field visualization
  primitive.
\newblock In {\em Visualization '99. Proceedings}, pages 263--532, 1999.

\bibitem{sujudi_identification_1995}
David Sujudi and Robert Haimes.
\newblock {\em Identification Of Swirling Flow In 3-D Vector Fields}.
\newblock 1995.

\bibitem{sprott_simple_1997}
J.~C. Sprott.
\newblock Some simple chaotic jerk functions.
\newblock {\em American Journal of Physics}, 65(6):537, 1997.

\bibitem{sprott_chaos_2003}
Julien~C. Sprott.
\newblock {\em Chaos and Time-series Analysis}.
\newblock Oxford University Press, January 2003.

\bibitem{sprott_elegant_2010}
Julien~C. Sprott.
\newblock {\em Elegant Chaos: Algebraically Simple Chaotic Flows}.
\newblock World Scientific, 2010.

\bibitem{henri_poincare_sur_1886}
Henri Poincare.
\newblock Sur les courbes definies par une equation differentielle.
\newblock {\em J. Math. Pures Appl.}, 2:151--217, 1886.

\bibitem{henri_poincare_sur_1885}
Henri Poincare.
\newblock Sur les courbes definies par une equation differentielle.
\newblock {\em J. Math. Pures Appl.}, 1:167--244, 1885.

\bibitem{henri_poincare_sur_1882}
Henri Poincare.
\newblock Sur les courbes definies par une equation differentielle.
\newblock {\em J. Math. Pures Appl.}, 8:251--296, 1882.

\bibitem{henri_poincare_sur_1881}
Henri Poincare.
\newblock Sur les courbes definies par une equation differentielle.
\newblock {\em J. Math. Pures Appl.}, 7:375--422, 1881.

\bibitem{perko_differential_2001}
Lawrence Perko.
\newblock {\em Differential Equations and Dynamical Systems}.
\newblock Springer, 2001.

\bibitem{ott_chaos_2002}
Edward Ott.
\newblock {\em Chaos in Dynamical Systems}.
\newblock Cambridge University Press, August 2002.

\bibitem{arnold_singularities_1985}
V.~Vladimir~Igorevich Arnol'd, S.~M. Gusein-Zade, and A.~N. Varchenko.
\newblock {\em Singularities of Differentiable Maps: Volume 1: The
  Classification of Critical Points Caustics, Wave Fronts}.
\newblock Springer, January 1985.

\bibitem{arnold_catastrophe_1992}
Vladimir~Igorevich Arnol'd.
\newblock {\em Catastrophe theory: with 87 figures}.
\newblock Springer-Verlag, Berlin [etc.], 1992.

\bibitem{gilmore_catastrophe_1993}
Robert Gilmore.
\newblock {\em Catastrophe Theory for Scientists and Engineers}.
\newblock Dover Publications, April 1993.

\bibitem{byrne_connecting_2013}
Greg Byrne, Robert Gilmore, and Juan Cebral.
\newblock Connecting curves in higher dimensions.
\newblock \url{http://arxiv.org/abs/1305.2960}, September 2013.

\bibitem{letellier_large-scale_2005}
Christophe Letellier, Tsvetelin~D. Tsankov, Greg Byrne, and Robert Gilmore.
\newblock Large-scale structural reorganization of strange attractors.
\newblock {\em Phys. Rev. E}, 72(2):026212, August 2005.

\bibitem{poston_catastrophe_1996}
T~Poston and Ian Stewart.
\newblock {\em Catastrophe theory and its applications}.
\newblock Dover Publications, Mineola, {N.Y.}, 1996.

\bibitem{gilmore_connecting_2010}
R.~Gilmore, Jean-Marc Ginoux, Timothy Jones, C.~Letellier, and U.~S. Freitas.
\newblock Connecting curves for dynamical systems.
\newblock {\em J. Phys. A: Math. Theor.}, 43(25):255101, June 2010.

\end{thebibliography}

\end{document}